\documentclass{epl}
\title{Dynamical transitions in incommensurate systems}
\shorttitle{Dynamical transitions}
\author{L.~Consoli\thanks{E-mail: consoli@sci.kun.nl} \and H.~Knops \and A.~Fasolino}
\shortauthor{L. Consoli \etal}
\institute{
Institute for Theoretical Physics - University of Nijmegen  \\
           Toernooiveld 1 - 6525 ED Nijmegen \\
           The Netherlands
}

\pacs{05.45.-a}{Nonlinear dynamics and nonlinear dynamical systems}
\pacs{64.60.Ht}{Dynamical critical phenomena}
\pacs{68.35.-p}{Solid surfaces and solid-solid interfaces: structure and energetics}

\begin{document}

\maketitle

\begin{abstract}
In the dynamics of the undamped Frenkel-Kontorova model with kinetic
terms, we find a transition  between two regimes, a floating
incommensurate and a pinned incommensurate phase. This behavior is compared to
the static version of the model. While
in the static case the two regimes are separated by a single transition (the Aubry 
transition), in 
the dynamical case the transition is characterized by a critical region, in which 
different phenomena take place at different times.  
In this paper, the generalized angular momentum we have previously introduced, and the
dynamical modulation function are used to begin a characterization  of this critical region.
We further elucidate the relation between these two quantities, and present preliminary
results about the order of the dynamical transition.
\end{abstract}

\section{Introduction}
The Frenkel-Kontorova (FK) model\cite{fk} describes the interaction of a harmonic chain of atoms 
with a rigid substrate with period incommensurate to the lattice parameter of the chain.
Its generality makes it a powerful model 
to investigate many different physical systems\cite{ads, jos}, and in particular
microscopic friction between contacting surfaces\cite{strunz, bra, Con_prl}.
The static version of the model is characterized by 
the Aubry transition\cite{trans}, from a floating to a pinned state, for
a critical value $\lambda_c$ of the substrate modulation potential.
Using the undamped dynamical version of this model, we addressed 
in previous papers the topic of ``dissipation'' (in the sense of transfer of energy from 
the center of mass to phonon modes) in incommensurate structures: we 
have studied the mechanism (parametric resonances) that governs the onset of sliding 
friction\cite{Con_prl}, and the conditions under which a 
new conserved quantity can be defined, which can be seen as a  Generalized Angular 
Momentum (GAM) in the complex plane\cite{Con_pre}. 

In this work, we present new results, showing that, in the dynamics, a   
floating-to-pinned transition,  analogous to the static Aubry transition, 
is found for all values of the potential $\lambda < \lambda_c$.
The transition is characterized by a region of critical times,
with a remarkably complex behavior.    
After describing 
the FK model, we show that the dynamical modulation function 
undergoes at a critical time $t_{c1}$ the same breaking of analiticity that, in the
static case, occurs at $\lambda_c$.
A second critical time $t_{c2}>t_{c1}$, at which the GAM conservation stops,
was identified in \cite{Con_pre}; here, we show
explicitly the connection between these two quantities, by showing analytically 
that the analiticity of the modulation function implies conservation of the GAM.
We also present some initial results about the order of this dynamical
transition.

\section{The static model}

\begin{figure}
\twoimages{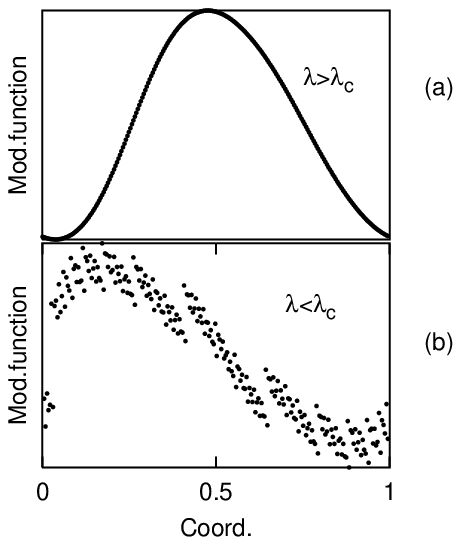}{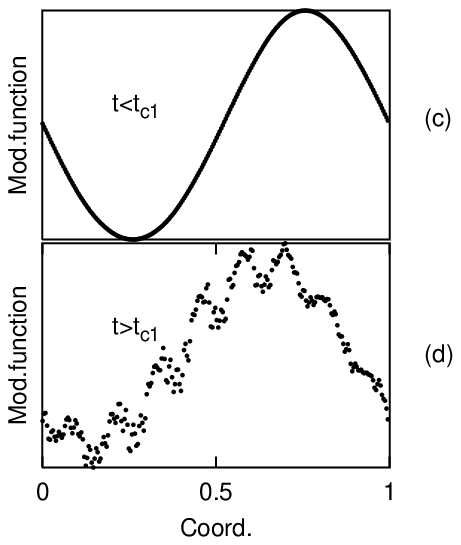}
\caption{Left: the modulation function in the static FK model. (a): for $\lambda < \lambda_c$ the 
function is analytic; (b) for $\lambda > \lambda_c$ it becomes non-analytic. Right: 
two snapshots of the dynamical modulation 
function. (c) $N=233$, $\lambda=0.05$, $t=300 < t_{c1}$ ; (d) $N=233$, $\lambda=0.05$, $t=820 > t_{c1}$. 
Even if $\lambda < \lambda_c$, the modulation 
function becomes non-analytic.}
\label{fig1}
\end{figure}

The FK model describes a chain of $N$ atoms in contact with a 
substrate, described by a rigid periodic potential. The 
intrachain interaction is modelled by a first neighbor 
harmonic potential, and the total potential energy reads:
\begin{equation}
{\cal V}=\sum_n\left[\frac{1}{2}\left(u_{n+1}-u_n-l
\right)^2 
+\frac{\lambda}{2\pi}\sin{ (\frac{2\pi u_n}{m} )}\right]
\label{v_stat}
\end{equation}
where $u_n$ are the lattice positions, $l$ is the average atomic distance, and 
$\lambda$ is the strength of the substrate potential rescaled to 
the spring constant.  The ratio between $l$ and the modulation potential period $m$ is chosen to
be an irrational number (in our case, $m=1$ and $l=\tau=(\sqrt{5}+1)/2$, the golden mean), in order to
simulate an incommensurate (IC) modulated structure. 
Numerically, we use a finite chain of length $Nl$ and use ratio of consecutive Fibonacci 
numbers to approximate incommensurability\cite{Con_prl}.
The Aubry transition can be monitored by the modulation function (see e.g. \cite{jos}). 
It is defined as: $f(nl\,\,mod\,\,m)=u_n - nl - Q$, where $Q=\frac{1}{N}\sum_n u_n$
is the center of mass (CM) of the chain, and  contains information 
about the groundstate positions of the chain particles with respect to the substrate.
When $\lambda>\lambda_c$, it becomes non-analytic, as shown in the left side of  fig.~\ref{fig1}. 
Another feature of the FK model is the existence of a zero frequency phason mode below
$\lambda_c$ that moves to finite frequency at the Aubry transition. 
If the interaction with the substrate is weak ($\lambda<<\lambda_c$), deviations 
from equidistant spacing $l$ in the groundstate are modulated with the 
substrate modulation wave-vector $q=2\pi l$, due to the frozen-in phonon $\omega_q$. 
For values of the coupling $\lambda < \lambda_c$, the Fourier transform of the displacement related
to $q$ has thus the strongest amplitude, and amplitudes of higher harmonics $nq$  
scale with $\lambda^{|n|}$.
It is therefore natural to use $|n|$ instead of $k$ and relabel the modes (see 
\cite{Con_pre} for the details of the relabeling procedure). In the left side of fig.~\ref{fig2}, it can 
be seen that, both with the usual labels (panel (a)) and with the relabeling (panel (b)),
the effect of the modulation is lost when $\lambda > \lambda_c$, which for this 
model is $\lambda_c=0.154...$.
 
\section{The dynamical model}

In the dynamical version
of the model, a kinetic term is added to the potential energy; in our case, it 
corresponds to giving an initial velocity to the particles of the chain. The
Hamiltonian now reads:
\begin{equation}
{\cal H} = {\cal V} + {\cal K} = {\cal V} + \sum_n\frac{p_n^2}{2}
\label{h_dyn}
\end{equation}
where the $p_n$ are the particle momenta. The CM velocity is defined as: $P=\frac{1}{N}\sum_n p_n$.
The dynamics of the system can be studied by following
the time evolution of the CM of the chain Q, and of the particles deviations $x_n = u_n - nl - Q$.
As initial conditions we choose $p_n(t=0)=P_0$ and $x_n(t=0)$ corresponding to the groundstate.
The equations of motion for the deviations are numerically solved with a Runge-Kutta-Fehlberg algorithm.
Convergence of the results is found for a tolerance of $10^{-9}$, ensuring total energy conservation up to
$10^{-8}$.
As we saw in the previous section, phonon amplitudes 
play a crucial role in the statics. In the dynamics, their behavior can be used to elucidate 
the time evolution of the system. The inherent nonlinear coupling of the CM motion to the phonons of the chain
leads to an irreversible decay of the CM velocity, that in an Hamiltonian 
picture can be identified with the onset of sliding
friction. Transfer of kinetic energy from the CM to the internal vibrations occurs via a complex sequence of parametric
resonances mediated by ordinary resonances with phonons related to the modulation potential\cite{Con_prl}. These excitations can 
further combine and give rise to the appearance 
of Umklapp terms. This phenomenon can be monitored by the breakdown of the conservation of 
the GAM, as derived in \cite{Con_pre}:
\begin{equation}
p_{\phi} = -i\sum_n nqa_{-nq}\dot{a}_{nq}
+\frac{q}{2\pi}\dot{Q} \equiv L + \frac{q}{2\pi}\dot{Q}
\label{gam} 
\end{equation}
where $a_{\pm nq}$ are the 
phonon amplitudes, obtained by Fourier transforming at each timestep the expression for the 
chain distorsions $x_n$. 

\section{Modulation function and phonon amplitudes}

We have shown in the previous section how the behavior of the modulation
function changes when going through the transition at $\lambda_c$.
If we now turn to the dynamical FK model with initial velocity, it is very 
intriguing to see that, for all values
$\lambda < \lambda_c$, the same behavior is found, after a critical time $t_{c1}$. 
\begin{figure}
\twoimages{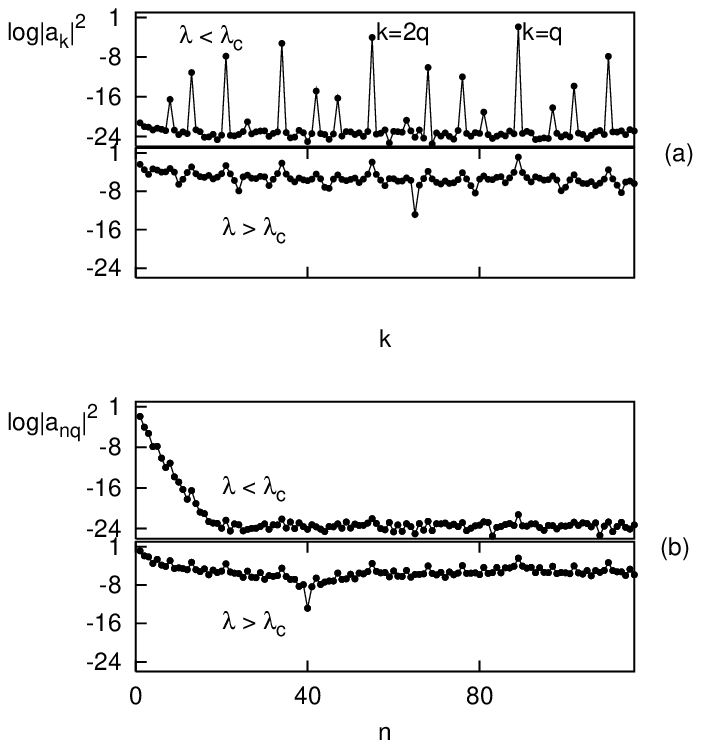}{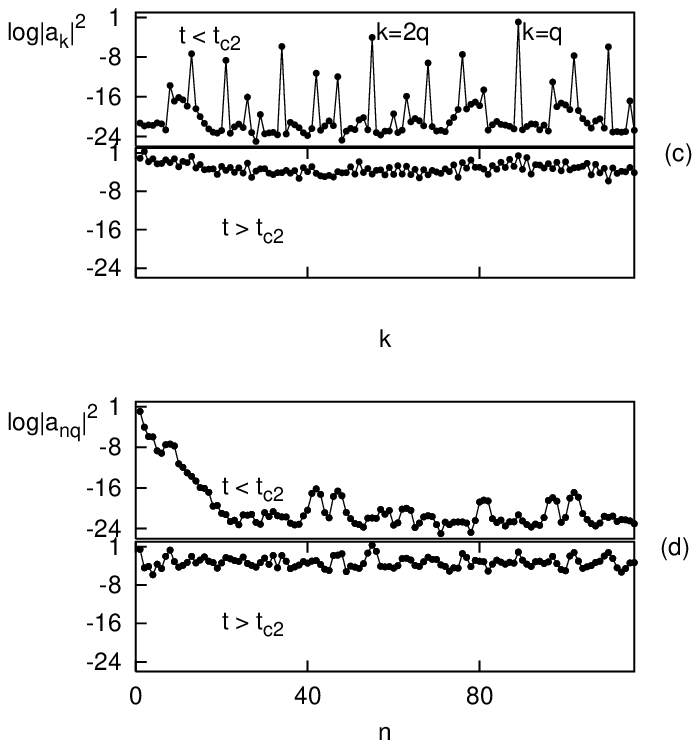}
\caption{(a) Phonon amplitudes in the static FK model, for the cases $N=233$, $\lambda=0.05 < \lambda_c$, 
and $\lambda=0.16 > \lambda_c$. The modes are labeled in the usual way. The first two modulation
wave-vectors $k=q,2q$ are shown. (b) Same, but the 
relabeling $k\rightarrow n$ is used  to make the scaling behavior apparent. (c) Phonon amplitudes in the 
dynamical case, for $N=233$, $P_0=0.29$, $\lambda=0.05 < \lambda_c$, $t=300 < t_c$, and $t=1000 > t_c$. Modes 
are labeled as in (a). (d) Same, with relabeling as in (b).}
\label{fig2}
\end{figure}

This can be shown by extending the concept of modulation function to the dynamics. Since the value 
$q=2\pi l$ is fixed, we do it in a straightforward way: $f_t (nl\,\,mod\,\,1)\equiv u_n(t) - nl - Q(t)\equiv x_n(t)$. 
The results are shown in panels (c) and (d) of fig.~\ref{fig1}. The similarity with 
the static case is apparent.
\begin{figure}
\twoimages{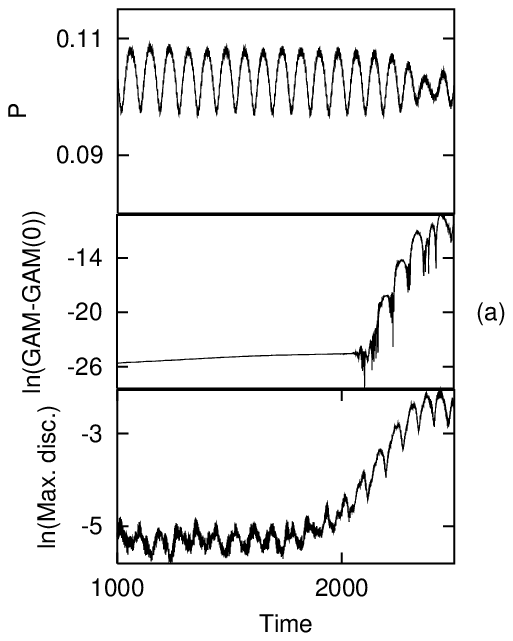}{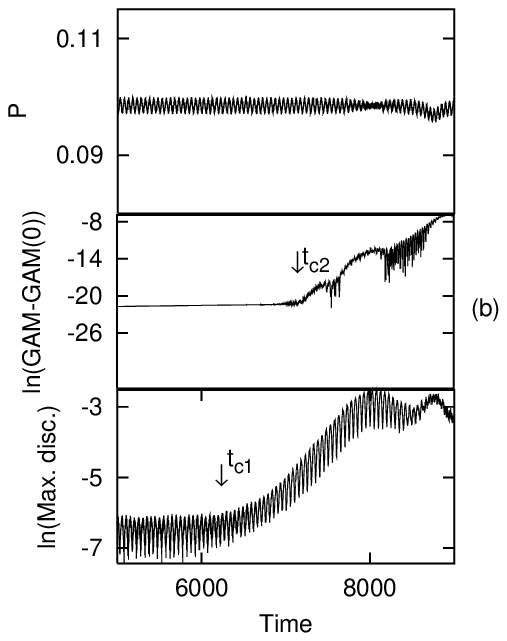}
\caption{Behavior of the deviations of the GAM from the initial value GAM(0) and of the largest discontinuity of the
modulation function. (a) $N=144$, $\lambda=0.05$, $P_0=0.11$. This velocity
corresponds to the resonance $2\pi P_0=\omega_{2q}/2$. (b) Same $N$ and $\lambda$, but with $P_0=0.1$. The 
two values of $P_0$ are very close,
but the behavior of the CM momentum is quite different since the second value is
not a resonance. Nevertheless, the GAM and the largest discontinuity have the same kind of
temporal evolution. This stresses the generality of the transition mechanism.}
\label{fig3}
\end{figure}
The time $t_{c1}$ can be accurately determined from the largest discontinuity of the modulation function (see fig.~\ref{fig3}, lower panels).
Another signature of the dynamical transition is that it is accompanied by a breakdown of the conservation of the GAM. This is shown in 
the middle panels of fig.~\ref{fig3}. It is however remarkable that this breakdown always occurs at a later time $t_{c2}>t_{c1}$. Fig.~\ref{fig4}
shows that this remains true for all system sizes considered. Qualitatively, one can understand this because the non-analiticity of the 
modulation function is caused by the excitation of so many phonon modes as to render the Fourier series representing it not absolutely convergent.
\begin{figure}
\onefigure{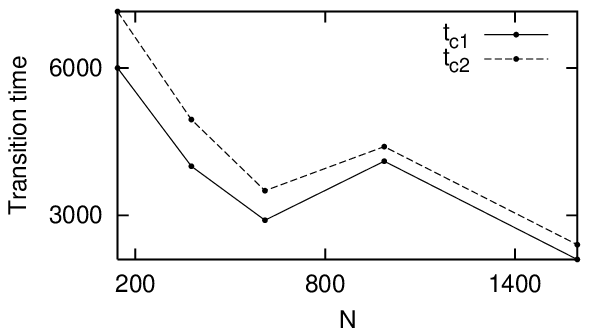}
\caption{Behavior of $t_{c1}$ and $t_{c2}$ as a function of the number of particles $N$, for 
$P_0=0.1$, $\lambda=0.05$, and $N=144,377,610,987,1597$.
For all cases, $t_{c2}>t_{c1}$.}
\label{fig4}
\end{figure} 
The conservation of the GAM only breaks down when this excitation reaches the zone boundary, generating Umklapp terms\cite{Con_pre}.
This is shown more formally in the next section. This dynamical transition can also be seen in the qualitative difference in 
the Fourier amplitudes shown in panels (c) and (d) of fig.~\ref{fig2}.

\section{The GAM and the modulation function}
\label{crit}

We show that analiticity of the modulation function implies conservation of the GAM.
The modulation function can be written as:
\begin{equation}
f(z) = \sum_m a_{mq} e^{i2\pi mz}
\label{hull}
\end{equation}
Indeed, writing the distortions $x_n$ as inverse Fourier transforms, with mode relabeling, gives:
\begin{equation}
x_n = \sum_m a_{mq}e^{inmq}
\label{dist}
\end{equation}
From eqs.~(\ref{hull}) and (\ref{dist}) it follows immediately that:
\begin{equation}
x_n = f(nl)
\label{hullcoord}
\end{equation}
This is true for the statics; taking into account the time dependence, we can generalize
to the dynamics: $a_m\rightarrow a_m(t)$, $f(z)\rightarrow f(z,t)$.
The equations of motion for the distorsions are\footnote{In eq.~(2) of ref.~\cite{Con_prl}
and eq.~(4) of ref.~\cite{Con_pre} the $\ddot{Q}$ term has been omitted.}:
\begin{equation}
\ddot{x}_{n} + \ddot{Q} = x_{n+1} + x_{n-1} -2x_n +\lambda\cos{(nq+2\pi x_n + 2\pi Q)}
\label{eom}
\end{equation}
From eqs.~(\ref{hullcoord}) and (\ref{eom}), we derive the equation of motion for $f$:
\begin{equation}
\ddot{f} = f(z + l,t) + f(z - l,t) -2f(z,t) - \ddot{Q}+\lambda\cos{(2\pi z + 2\pi f(z,t) + 2\pi Q(t))}
\label{hulleom}
\end{equation}
where we make the association $x_{n\pm 1} = z \pm l$.
Let us now take the first of the two terms that make up the GAM in eq.~(\ref{gam}) and rewrite it in terms
of the modulation function. If the modulation function $f$ is analytic, we can express the 
coefficients $a_{nq}$ as:$a_{nq} = \int_0^{1}dz e^{-in2\pi z}f(z,t)$ and using partial 
integration, we obtain:
\begin{equation}
L = -i\sum_n nqa_{-nq}\dot{a}_{nq} = \frac{q}{2\pi}\int_0^1 dz\frac{df(z,t)}{dz}\dot{f}(z,t)
\label{lhull}
\end{equation}
We now compute the first time derivative of eq.~(\ref{lhull})
It can be shown that, due to periodic integration boundaries, only one term remains:
\begin{equation}
\dot{L} = \frac{q}{2\pi}\lambda\int_0^1 dz\frac{df(z,t)}{dz}\cos{(2\pi(z + f(z,t) + Q(t))}
\label{step}
\end{equation}
Using $f+z$ as a new integration variable and periodic integration boundaries, we get: 
\begin{equation}
\dot{L} = -\frac{q}{2\pi}\lambda\int_0^1 dz \cos{(2\pi(f(z,t) + z + Q(t))}
\label{quasi}
\end{equation}
From eq.~(\ref{eom}), we can now obtain the 
expression for the equation of motion of the CM $Q$:
\begin{equation}
\ddot{Q} = \frac{\lambda}{N}\sum_n \cos{(nq + 2\pi x_n + 2\pi Q)}
\label{qsum}
\end{equation} 
Under the assumption that we made of the analiticity of $f$, we can go from the summation to the integral and,
using eq.~(\ref{hulleom}), we obtain:
\begin{equation}
\ddot{Q} = \lambda\int d\phi \cos{(2\pi(z + f(z,t) + Q(t))}
\label{qint}
\end{equation}
Comparison between eqs.~(\ref{quasi}) and (\ref{qint}) gives us (after integration):
\begin{equation}
L + \frac{q}{2\pi}\dot{Q} = const
\label{fine}
\end{equation}
i.e., we have derived the conservation of the GAM from the definition of the modulation function.

The crucial assumption in the calculation is the analiticity of the modulation function, which in the static
case defines the regime under the Aubry transition. Notice that this result is compatible with our
observation that $t_{c2}>t_{c1}$: analiticity of the modulation function implies GAM conservation, but the contrary is not
necessary. Lastly, we notice from fig.~\ref{fig3} that this 
mechanism is robust: while resonances affect in a significant way the CM velocity, the GAM and the 
modulation function retain the same kind of behavior.

\section{Order of the transitions and relation with the static case}
\label{order}

\begin{figure}
\twoimages{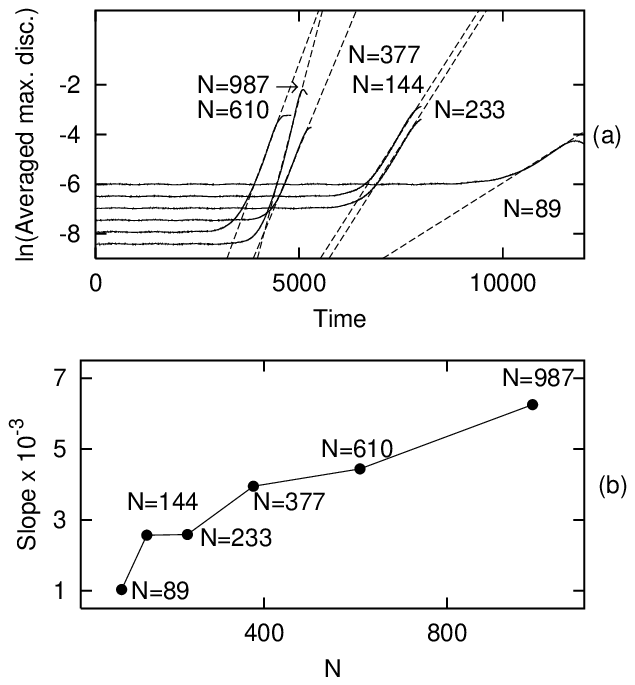}{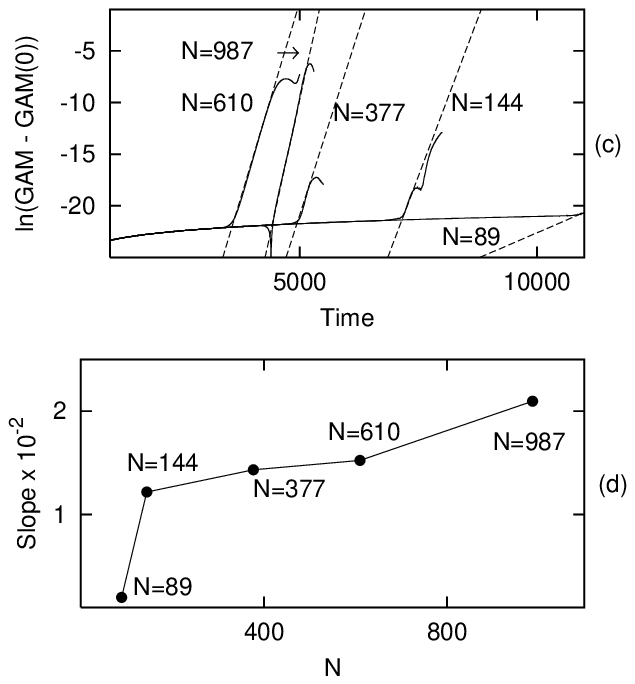}
\caption{(a) Exponential fits for the growth of the largest discontinuity in the modulation function for various $N$ at $\lambda=0.05$,
$P_0=0.10$. The initial values are determines by numerical accuracy. (b) Behavior of the fitting exponents shown in (a); the increase
with $N$ is apparent. (c) Exponential fits for the GAM. (d) Exponents of the GAM fit as function of $N$; here too, they 
increase with $N$.} 
\label{fig5}
\end{figure}

Both transitions we observe in the dynamical case are very sharp; this suggests that, in the limit of an infinite
chain, they could be first-order. This would be in contrast with the static case, where it
has been proven (see e.g. \cite{jos}) that the largest discontinuity in the modulation functions undergoes a second 
order transition. 
Presently, we do not have the analytical tools to check this rigorously. We can however get some 
numerical indications: we fit the behavior of the GAM and of the largest
discontinuity to an exponential function, and study the behavior of the 
exponent as a function of the size $N$. In fig.~\ref{fig5}, we see that, in both cases, the value 
of the fitting exponent grows when $N$ increases.
This indicates that, for $N\rightarrow\infty$, the exponent
could diverge and the transition would become first-order. More work is needed to confirm these initial results.   
It is also possible to relate in a direct way the dynamical and static behavior. 
\begin{figure}
\onefigure{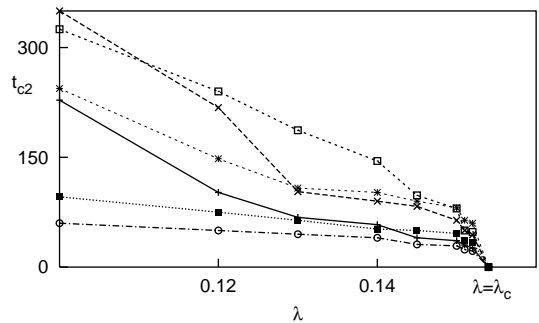}
\caption{Behavior of the transition time $t_{c2}$ as a function of the substrate potential $\lambda$ in the 
proximity of $\lambda_c=0.154...$.
Each line corresponds to a different initial velocity. $t_{c2}\rightarrow 0$ 
for $\lambda\rightarrow \lambda_c$.}
\label{fig6}
\end{figure}
This is shown in fig.~\ref{fig6} for $t_{c2}$: when 
the substrate potential approaches the critical static value $\lambda_c$, $t_{c2}$ goes to zero.
This allows us to relate the statics and the dynamics of the FK model.

\section{Conclusions}
\label{concl}

We presented in this paper numerical and analytical evidence that, in the kinetic FK model, a region is found where
dynamical transitions take place. This region 
separates a floating IC from a pinned IC phase, and is therefore equivalent to the static Aubry transition
(and, in fact, reducing to it in the limit of $\lambda\rightarrow\lambda_c$). Non-analiticity of the modulation
function and breakdown of conservation of the GAM characterize these transitions.
More work is needed to assess the order of these transitions and the relationship between 
$t_{c1}$ and $t_{c2}$, investigating for example the possibility that, for 
$N\rightarrow\infty$, $t_{c1}$ and $t_{c2}$ converge to a single value $t_c$.

\acknowledgments

We would like to thank Ted Janssen for productive discussions.

\end{document}